# A Purity Monitoring System
# for the H1 Liquid Argon Calorimeter

**Abstract**

The ionization probes used for monitoring the liquid argon purity in the H1 calorimeter are described and results of their operation in tests at CERN and during the period 1992 to the end of 1998 at HERA are given. The high sensitivity of the charge measurements leads to refined charge collection models, and to the observation of a variation of the ionization yield of our electron sources with temperature.

*To be submitted to Nuclear Instruments and Methods*

E. Barrelet[*,12], B. Andrieu[11], A. Babaev[8], E. Banas[2], D. Bederede[4], P. Biddulph[5,16],
K. Borras[3,17], F. Brasse[5], V. Brisson[10], P. Burmeister[5], G. Buschhorn[9], B. Canton[12],
U. Cornett[5], G. Cozzika[4], J. Cvach[13], A. Cyz[2], D. Darvill[5], M. David[4,†],
F. Descamps[12,18], A. Drescher[3,19], U. Dretzler[3], G. Ernst[3], G. Falley[5], R. Felst[5],
J. Feltesse[4], J. Ferencei[6,17], W. Flauger[5,†], M. Fleischer[3,17], J. Formánek[14], K. Gadow[5],
J. Gayler[5], E. Gažo[6], J.-F. Genat[12], J. Godlewski[2], Ch. Grégory[11], M. Grewe[3],
G. Grindhammer[9], L. Hajduk[2], M. Hensel[3,17], I. Herynek[13], J. Hladký[13], D. Imbault[12],
Ç. İşsever[3], S. Karstensen[5], C. Kiesling[9], M. Kolander[3], J. Koll[5], V. Korbel[5], F. Krivan[6],
H. Küster[5], R. Lemrani[5], P. Loch[5], J.-P. Lottin[4], J. Marks[5], F. Martin[12], J. Martyniak[2],
S. Mikocki[2], F. Moreau[11], J. Naumann[3], C. Niebuhr[5], G. Nowak[2], H. Oberlack[9],
P. Pailler[4], K. Rauschnabel[3,21], P. Reimer[13], F. Rossel[12], P. Schacht[9], J. Schaffran[5],
H. Schmücker[9], V. Schwatchkin[8], Y. Sirois[11], J. Spalek[6], J. Spiekermann[3,22]
H. Steiner[12,23], U. Straumann[15], M. Sueur[4], K. Thiele[5], N. Tobien[5], W. Tribanek[9],
S. Valkár[14], C. Vallée[7], G. Villet[4], D. Wegener[3], T. P. Yiou[12] and J. Žáček[14]

* Corresponding author. *E-mail address : barrelet@in2p3.fr*
[1] *III. Physikalisches Institut der RWTH, Aachen, Germany[a]*
[2] *Institute for Nuclear Physics, Cracow, Poland[b]*
[3] *Institut für Physik, Universität Dortmund, Dortmund, Germany[a]*
[4] *CEA, DSM/DAPNIA, CE-Saclay, Gif-sur-Yvette, France*
[5] *DESY, Hamburg, Germany*
[6] *Institute of Experimental Physics, Slovak Academy of Sciences, Košice, Slovak Republic[c,d]*
[7] *CPPM, CNRS/IN2P3 - Université Méditerranée, Marseille, France*
[8] *Institute for Theoretical and Experimental Physics, Moscow, Russia[e]*
[9] *Max-Planck-Institut für Physik, München, Germany*
[10] *LAL, Université de Paris-Sud, IN2P3-CNRS, Orsay, France*
[11] *LPNHE, Ecole Polytechnique, IN2P3-CNRS, Palaiseau, France*
[12] *LPNHE, Universités Paris VI and VII, IN2P3-CNRS, Paris, France*
[13] *Institute of Physics, Czech Academy of Sciences, Praha, Czech Republic[c,f]*
[14] *Faculty of Mathematics and Physics, Charles University, Praha, Czech Republic[c,f]*
[15] *Physik-Institut der Universität Zürich, Zürich, Switzerland[g]*
[16] *Now at Department of Physics and Astronomy, University of Manchester, Manchester, UK*
[17] *Now at DESY*
[18] *Now at Institut Max Von Laue - Paul Langevin, Grenoble, France*
[19] *Now at Telekom, Bonn, Germany*
[20] *Now at Physikalisches Institut, Universität Heidelberg, Heidelberg, Germany*
[21] *Now at FH Esslingen, Germany*
[22] *Now at Software Design Management, München, Germany*
[23] *Department of Physics and LBL, University of California, Berkeley, USA*
[a] *Supported by the Bundesministerium für Bildung und Forschung, FRG, under contract numbers 05 H1 1PAB /9 and 05 H1 1PEA/6*
[b] *Partially supported by the Polish State Committee for Scientific Research, grant no. 2P0310318 and SPUB/DESY/P03/DZ-1/99, and by the German Bundesministerium für Bildung und Forschung, FRG*
[c] *Supported by the Deutsche Forschungsgemeinschaft*
[d] *Supported by VEGA SR grant no. 2/5167/98*
[e] *Partially Supported by Russian Foundation for Basic Research, grant no. 00-15-96584*
[f] *Supported by GA AV ČR grant no. A1010821*
[g] *Supported by the Swiss National Science Foundation*
† *Deceased*

## Introduction

The argon purity monitoring system described in this paper has been successfully operating in the liquid argon calorimeter of the H1 detector at the HERA *e-p* collider since 1992. The detector is used to make detailed studies of the final states produced when 27.5 GeV electrons or positrons collide with 920 GeV protons, and to determine the structure function $F_2(x, Q^2)$ of the proton in kinematical regions which have heretofore been inaccessible. To this end a large sampling calorimeter consisting of lead and steel plates immersed in liquid argon and covering a large fraction of the total available solid angle is used to determine the energies and angles of the outgoing photons, electrons and hadrons. To realize the experimental goals an energy resolution for electrons of $\sigma(E)/E = 12\%/\sqrt{E}$ and a stability of the response at the 1 % level are required. We refer the reader to references[1][2] for a detailed description of this detector.

In this paper we confine ourselves to a description of the purity monitoring system which is being used to meet these objectives. Even small concentrations of electronegative impurities such as oxygen or halogens in liquid argon (LAr) at levels of less than 1 part per million (ppm) can seriously impair the electron collection efficiency and thus the energy resolution of the calorimeter. It is therefore necessary to measure and monitor either the presence of the impurities directly, or even better, the electron collection efficiency during the actual operation. We have chosen the latter option using ten small ionization probes situated at strategic locations in the detector. With these probes it is possible to monitor the effect of impurities on the calorimeter performance at the 0.1 % level. The information given by the probes can be described in terms of the electron mean free path $\lambda_e$ in the active medium (LAr) of the calorimeter. If this mean free path can be made large compared to the typical drift length of the electrons produced in the primary ionization process, the collection efficiency will be high and the output signals will be directly proportional to the energy deposited in the calorimeter. On the other hand, if some of the electrons are swallowed by impurities, the calorimeter response will be smaller in amplitude and worse in resolution. The monitoring system has worked extremely well, not only to detect impurities, but also to detect subtle changes in detector performance such as for example gain variations with temperature of the LAr, that otherwise would almost certainly have escaped notice.

This paper is divided into five sections. In section 1 we describe the probes and the associated electronics. Section 2 defines the methods used to analyze the data. Section 3 deals with the unanticipated LAr temperature effect. Section 4 is devoted to the absolute calibration of the probes using variable fields and drift distances. Finally, in Section 5 we discuss the initial use of the probes in tests of the calorimeter modules at CERN and their performance in the H1 calorimeter up to the present time.

## 1 Description

The general features of the liquid argon monitoring system and some results have been given in a previous paper [2]. In this chapter the whole system will be presented in detail.

### 1.1 Description of the probes

The structure of the probes is presented schematically in figure 1.

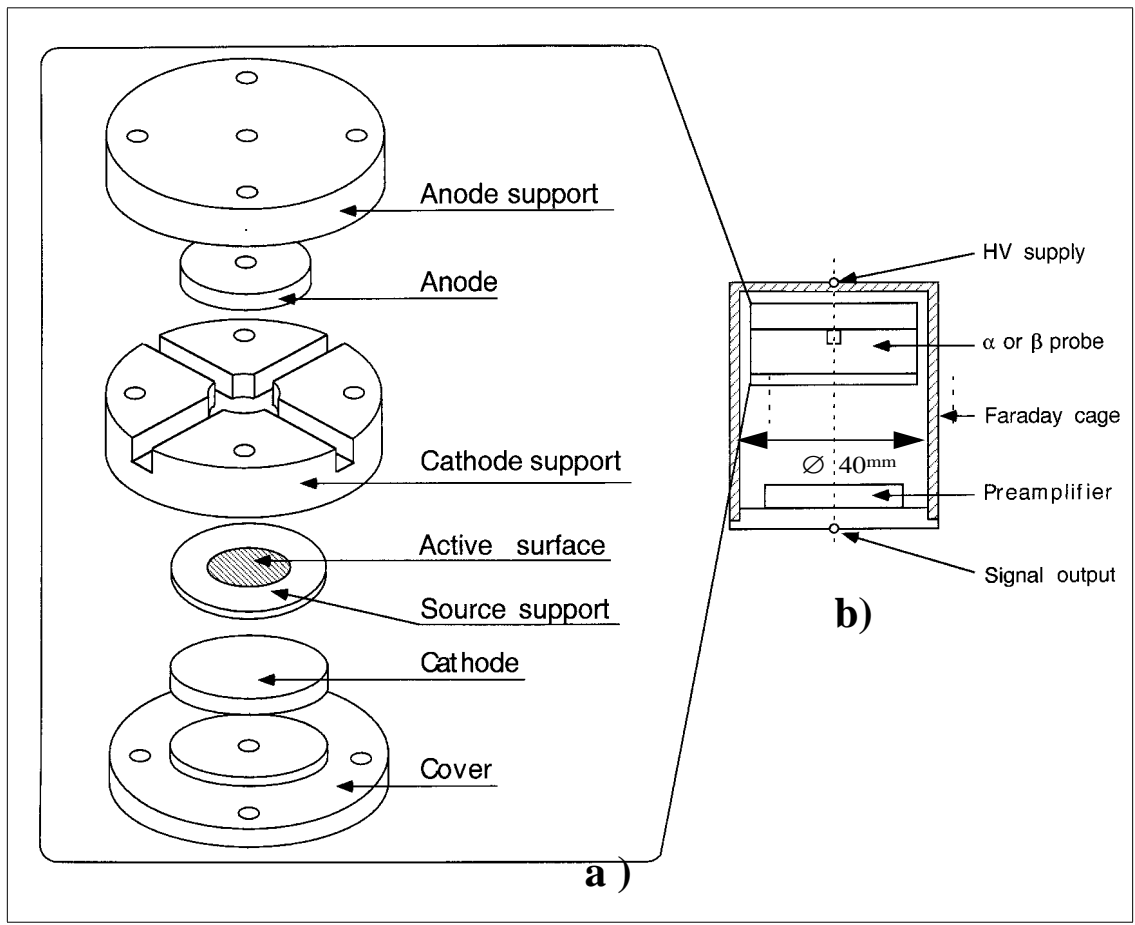

**Figure 1:** Schematic view of the LAr purity probes: a) Elements of the ionization chamber, b) Structure of the probes.

There are three main components:

- A small ionization chamber which is based on a design kindly supplied by D.C.Rahm. The different elements of the chamber are shown in figure 1a. The anode and the cathode are made of aluminum sheet whereas their supports and the spacer defining the liquid argon gap thickness are made of Teflon (PTFE).

- A radioactive source, either the α emitter [241]Am (Amersham International Plc, AMR 23, 7.4 kBq) or the internal conversion electron emitter [207]Bi (Isotope Products Laboratory, ME-207, 34 kBq), is used to ionize the LAr gap.
  As the "practical range" in liquid argon is quite different for an [241]Am α particle (≈ 35 μm) and for a [207]Bi electron (≈4.5 mm[1]), the gap thickness is chosen accordingly. It is 2 mm for [241]Am and 4 mm or 6 mm for [207]Bi. Bismuth is the only available monoenergetic electron source with a sufficiently long half-life (30.2 years), but it poses a problem because of its large electron range. For this reason it was proposed in [3] to include a "Frisch grid" in the ionization chamber. However from an analysis detailed in [4], we concluded that a grid makes the range problem even worse and we used instead the spectrum analysis outlined in § 2.2.

- Local electronics consisting of a low noise preamplifier, a neon lamp protecting it against sparking and a high quality glass capacitor (≈10 pF) for calibration.

---

[1] this number is justified in § 2.2.2

The whole assembly is housed within a cylindrical brass case with holes that allow the liquid argon to circulate.

## 1.2 Location of the probes in the cryostat

The purity monitoring system used for the H1 LAr calorimeter was developed during the tests of individual calorimeter modules in the years 1989 to 1992 at CERN. The early system at CERN contained only two [207]Bi and two [241]Am probes. In 1992, during the last two running periods, two [207]Bi probes were added. For each run the location of probes was adapted to the geometry of the calorimeter modules under test.

In 1991 two [241]Am and seven [207]Bi probes (respectively numbered 1,4 and 2,3,5,7,9,10,11) were installed in the H1 cryostat at DESY. Six [207]Bi probes with 4 mm gaps, are located at two central and four peripheral positions. In addition the interesting position near the expansion vessel (top right in figure 2), where the LAr moves due to convection currents, is equipped redundantly with [241]Am and both types of [207]Bi probes (respectively probe 1, probe 2 with a 6 mm gap and probe 3 with a 4 mm gap).

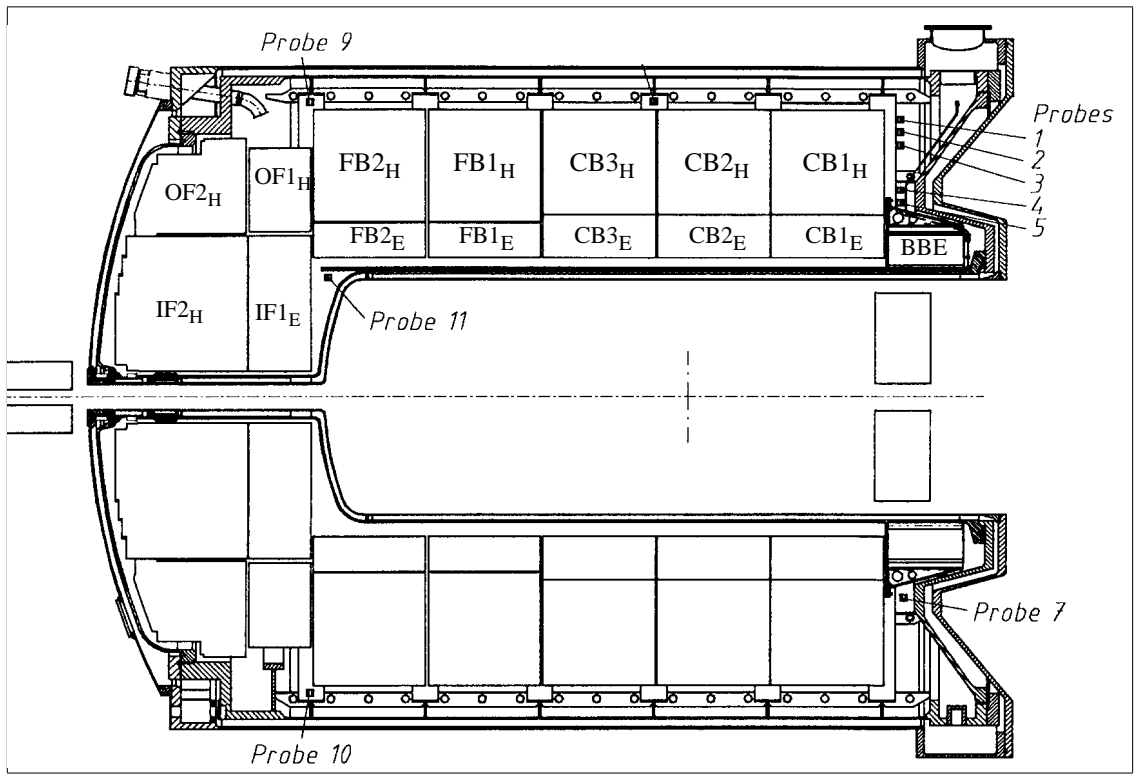

**Figure 2:** Location of purity probes inside the H1 LAr cryostat.

## 1.3 The electronic chain

The maximum electric charge collected by a purity probe is 4 orders of magnitude smaller than that produced by an electromagnetic shower in a typical H1 calorimeter channel, but the noise sources are the same. For this reason we have used a specific amplification chain (cf. fig.3), acting in 3 steps:

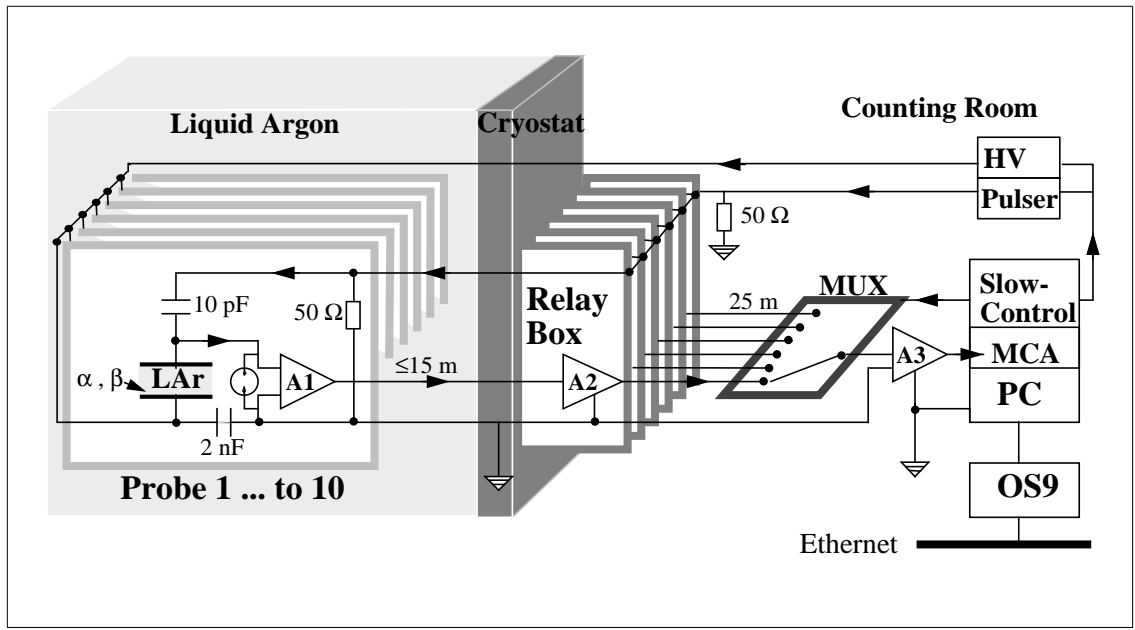

**Figure 3:** General layout of the purity monitor system.

1. A charge preamplifier A1, mounted inside each probe, which works at liquid argon temperature. It has been designed and used by experiment NA34 at CERN[5]. Its output response has a decay time of 60 µs. It is sent by a 50 Ω cable (≤ 15 m), traversing the LAr vessels via a standard "H1 feed-through", up to a "relay box" situated on the top of the H1 calorimeter cryostat.

2. A DC amplifier A2, in the relay box, which provide a supplementary gain of 10 in order to reduce the effect of pick-up noise occurring in the cable joining the relay box to the counting room.

3. An amplifier A3 in the counting room which is composed of:
   a) a differential receiver, integrated into the multiplexer (MUX), designed especially to reject the 50 Hz noise occurring between the ground potential of the cryostat and of the counting room.
   b) a commercial spectroscopy amplifier (*Ortec 571*) is used to give the signal a Gaussian shape peaked at 3 µs after pole-zero compensation. It provides a gain of 300.

Finally, the pulse height spectrum is made by a Multi Channel Analyzer (MCA) contained in a single PC add-on card (*Nucleus, PCA II*).

## 1.4 The calibration

The gain of the electronic chain is monitored by injecting an electronic pulse at the input of the charge preamplifier A1 during the accumulation of a spectrum. The standard calibration pulse height is adjusted in order to yield a peak above the highest energy produced by any radioactive source. The 10 probes are pulsed in parallel by the same pulser, at a 70 Hz rate which is 2% of the observable radioactive decay rate. In addition, the same pulse is injected at the input of an eleventh amplifier A2 in the relay box. This allows the calibration of the whole electronic chain (except amplifier A1) during each data acquisition cycle. For example, after recording the spectra of the 10 probes, one records systematically 2 spectra of this dummy "11th probe", respectively with a standard calibration pulse and a pulse 10 times smaller. By interpolation to a null pulse height, we monitor the value of the offset of the electronic chain, common to the 10 energy spectra but not directly measured.

The calibration pulse height is tuned by a voltage generated by a digital-to-analog converter situated in the slow-control card (*Keithley 570*) which charges a capacitor in the pulser (*ORTEC 419*) and then discharges it through a mercury relay and a dividing bridge (adapted to 50 Ω on both sides). The resulting pulse, presenting a 0.5 mV step, is received by a 10 pF capacitor inside each probe. The dividing bridge in the relay box attenuates in the same proportion the calibration voltage (from a few volts to 0.5 mV) and the low frequency noise between the cryostat and the counting room grounds. The pulser frequency is 70 Hz, in order not to be synchronous with the 50 Hz AC power.

The stability of the calibration system over 8 years of operation is excellent, better or equal to 0.1%. Its absolute precision is limited to 5-10%, due to our ignorance of the value of the resistors and capacitances at liquid argon temperature. The largest uncertainties comes from the resistance of the copper cables inside the cryostat and the parasitic capacitance in parallel to the 10 pF reference capacitor. The best absolute calibration comes from the study of the $^{207}$Bi spectrum. For instance the energy scale is determined by extrapolating at high electrical field the peak position measured for the 481.7 keV K-electrons.

## 1.5   The data acquisition

The core of the data acquisition system is a PC containing 2 add-on cards: the MCA card (cf. § 1.3) and the slow-control card (cf. § 1.4).

The control program has been written in C language (Borland TurboC). It incorporates the *Nucleus* library for the MCA card. All the parameters controlling the acquisition of events are downloaded in the MCA at initialization. With such a setup, the MCA accumulates an energy spectrum in 5 minutes (corrected for dead time) with a 2048 channel energy resolution, and it provides at the same time a live visualization of the spectrum on the PC screen. During the accumulation of a spectrum, the PC uses the digital voltmeter[2] function of the slow-control card to monitor repeatedly 32 channels, including all relevant HV or LV power supplies, calibration voltages, ground to ground AC or DC offsets etc. The 32 mean values and sigmas are systematically stored with each spectrum and sent (via a RS232 link) to the VME processor (OS9 system) in charge of the slow-control of the H1 calorimeters. Then the PC selects the next multiplexer channel and changes if necessary the reference level for the calibration pulses or the high voltage sent to the next purity probe (using the other functions of the slow-control card). The PC itself has been changed several times, without affecting the quality of the MCA and the slow-control cards.

The OS9 station performs an analysis of the energy spectrum, estimating mean energy, width and yield of various peaks (including calibration) as described hereafter. The value of these estimators and the slow-control measurements are kept in a data-base, while only a few spectra are stored from time to time for off-line controls.

---

[2]   the true 4 digit precision of the Keithley DVM has proved to be essential, in particular for monitoring the HV and the calibration reference levels.

## 2 Analysis of $^{241}$Am and $^{207}$Bi spectra

This section describes the first step of the purity analysis, i.e. the extraction of *the electron mean free path* $\lambda_e$ from the spectrum of charge collected in a probe. As this operation is much more complicated for the $^{207}$Bi electron spectrum than for the $^{241}$Am $\alpha$ spectrum, let us first explain why we need both types of sources:

The first advantage of an electron source is to be directly comparable to an electromagnetic shower inside the H1 calorimeter (same range of specific ionisation and same field strength). The second advantage discussed in § 4, is to give access to the absolute value of $\lambda_e$ by fitting the relative variation of $\lambda_e$ with field strength and/or by comparing the spectra obtained with two different absorption lengths (4 and 6 mm gap). However the $\alpha$ sources provide essential comparative information in the analysis of the temperature dependence of electron sources (see § 3).

### 2.1 The ionization spectrum of $^{241}$Am

The $^{241}$Am spectrum shows a single isolated asymmetric peak (2 unresolvable lines at $E_\alpha$=5.442 and 5.484 MeV). It is parametrized by a mean charge $Q_\alpha$ and an effective width $\sigma_\alpha$ (R.M.S.).

We operate at 20 kV/cm in order to minimize recombination losses. Nevertheless the non-recombined charge $Q_\alpha$ is just a quarter of the total ionization charge and it varies proportionally to the electrical field. The statistical precision of the $Q_\alpha$ measurement is 0.02% RMS. Because we are not able to separate recombination and attachment losses, we monitor the electrical field (constant at the 0.025% level) and measure the relative variation of $\lambda_e$ through the variation of $Q_\alpha$.

### 2.2 The ionization spectrum of $^{207}$Bi

#### 2.2.1 Experimental data

In figure 4 we show eight spectra taken at 2 day intervals during the test of an H1 calorimeter module in a CERN beam from April 25th to May 9th 1990. The impurities increased rapidly, allowing us to calibrate our device over a range of $\lambda_e$ larger than that seen later in H1 from 1991 to 1999. Looking at the sequence of spectra, both electron conversion lines are shifting to the left with time because $\lambda_e$ decreases with rising impurity, while the electronic calibration peak stays constant. Moreover, the response of the probes could be cross checked with the response to beam particles in the H1 calorimeter module in the same cryostat during the same period.

Bismuth emits $\gamma$s with three energies: $E_{\gamma_1} = 0.569 MeV$ ; $E_{\gamma_2} = 1.063 MeV$ ; $E_{\gamma_3} = 1.77 MeV$ . The splitting of the $\gamma_1$ peak into K(482 keV) and L(554 keV) electron lines is marked by a shoulder on the $\gamma_1$ falling edge. The splitting of the $\gamma_2$ peak is not resolved. Here the peak is wider because the range of the corresponding electrons is not negligible compared to the 6 mm gap. The continuous background is due to the Compton scattering for all 3 $\gamma$ lines.

#### 2.2.2 Numerical model for the $^{207}$Bi spectra

In order to monitor the composition of $^{207}$Bi spectra, we have defined an analytical function $S_{Bi}(Q)$ given in[4], where $Q$ is the measured charge, by adding the individual spectra corresponding to the seven $^{207}$Bi decay modes shown in figure 5a ($\gamma_1^K$, $\gamma_1^L$, $\gamma_1^{Compton}$..., $\gamma_3^{Compton}$). Each individual spectrum is computed by integration of the collected charge over all the phase space of the corresponding decay channel, assuming uniform ionization density along each electron track.

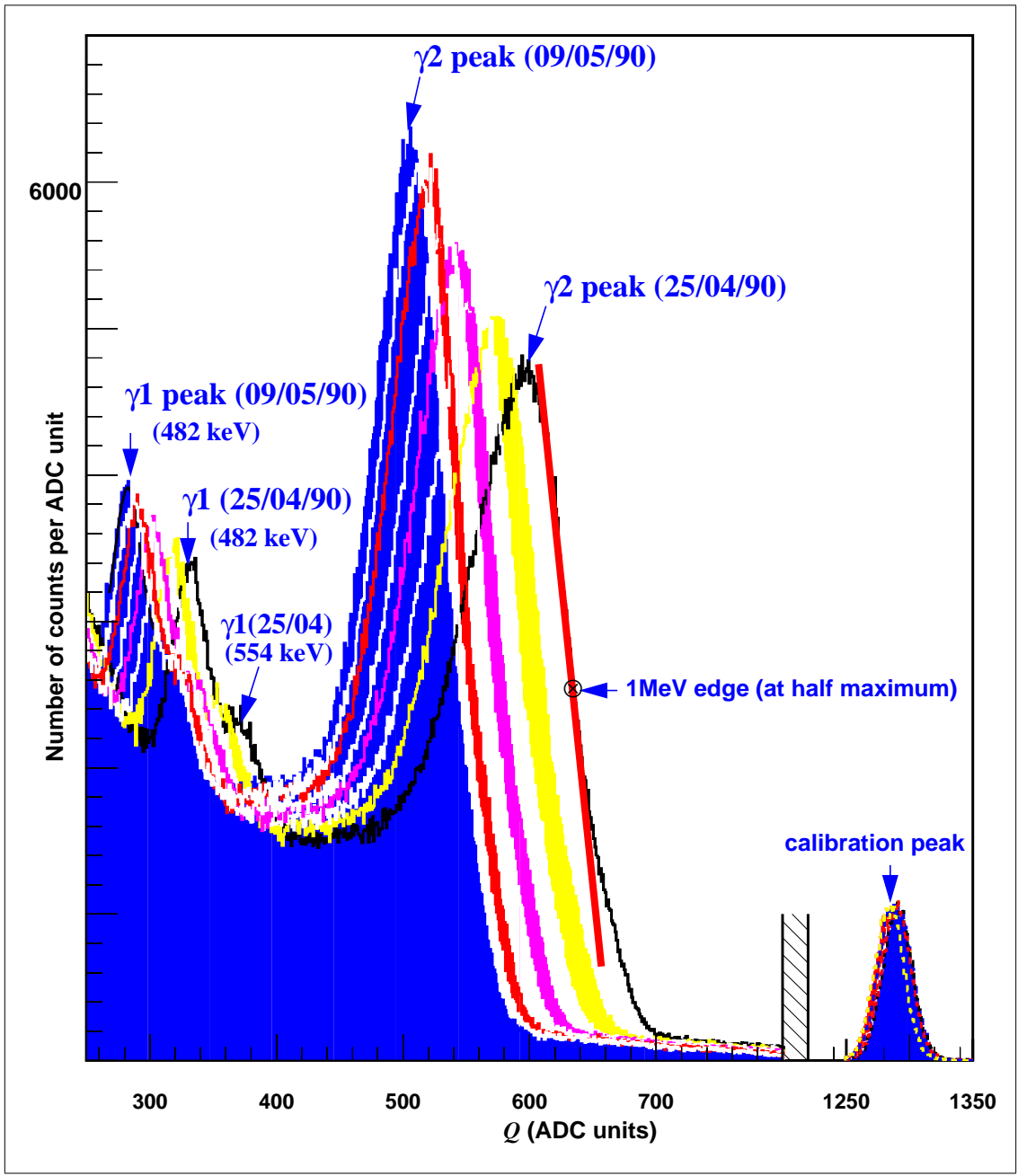

**Figure 4:** Eight $^{207}$Bi spectra accumulated during a single run at CERN (April-May 90). The electron conversion lines are shifted to the left with time due to the increasing impurities.

This model fits well to the eight $^{207}$Bi spectra taken with a 6 mm LAr gap (shown in figure 4) and eight others taken with a 4 mm gap (not shown). By changing only the parameter $\lambda_e$ from 4.32 mm to 3.14 mm, we reproduce the spectra taken on the first day of the CERN experiment and on the last (see figure 5 a and b). Similarly by changing the LAr gap width $D$ from 6 mm to 4 mm, we get good fits to the spectra taken with small gap probes.

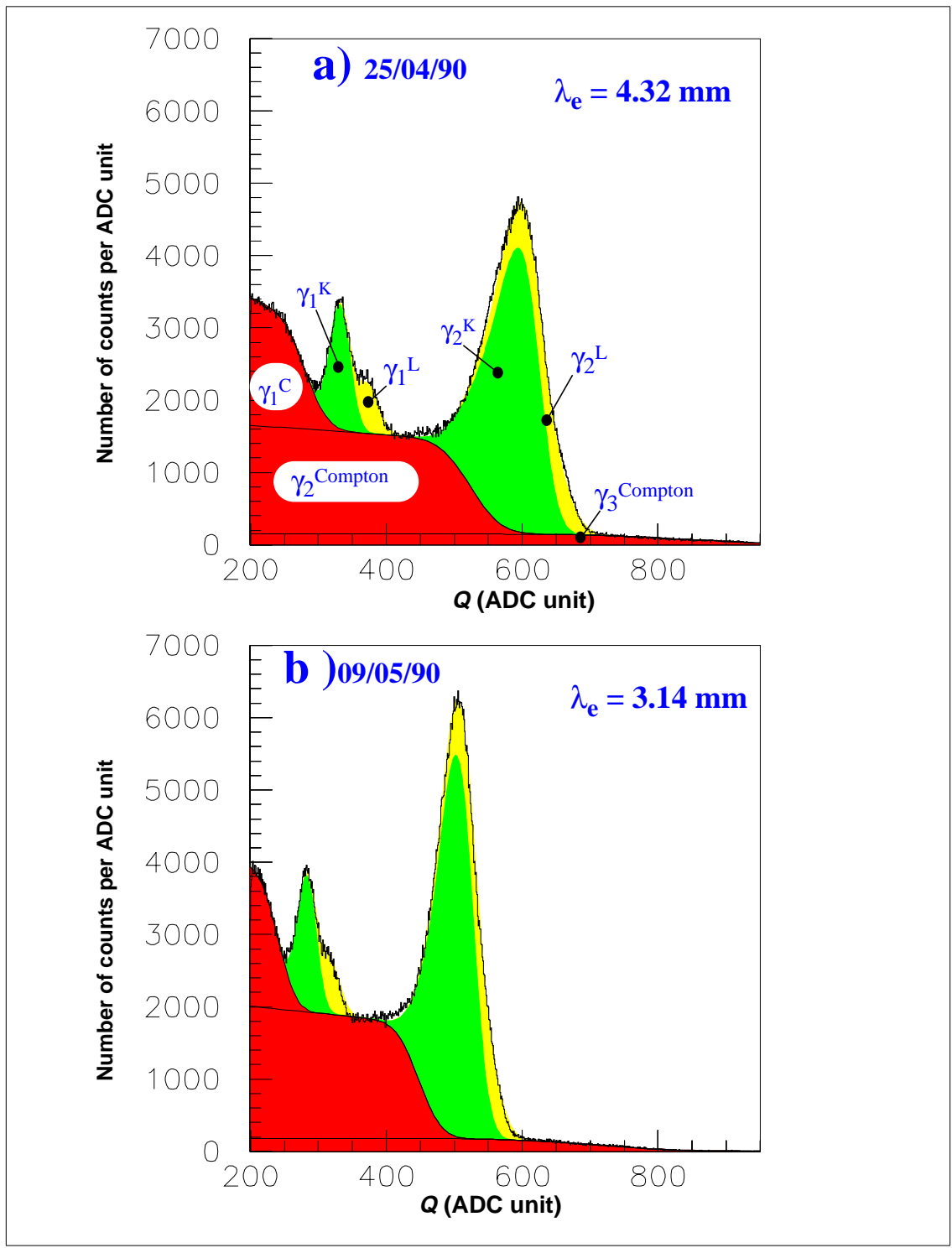

**Figure 5:** Comparison of the analytical function $S_{Bi}(Q)$ with spectra of a $^{207}$Bi probe with 6 mm gap. a) data of 25/04/90, b) data of 09/05/90. Areas corresponding to the three electron conversion mechanisms (K, L, Compton) are filled with different shades.

There are further parameters entering the analytical function $S_{Bi}(Q)$. Two parameters characterize each probe individually: the gain (ADC unit per collected charge) and the activity of the probe (number of decays per second). They are determined accurately from the first spectrum obtained for both gap sizes (at 25/04/90). The other parameters, common to all spectra, have been determined using the first spectrum of the 6 mm probe. First there are 7 branching ratios (2 electron conversion lines for $\gamma_1$, 2 for $\gamma_2$ and a Compton background for the 3 $\gamma$s). Second there are 2 form factors for each Compton background: the characteristic energy and the sharpness of the Compton edge. Third there is the range of the 1 MeV electrons in argon $R = 4.5 \pm 0.2$ mm, which is determined from the width of the $\gamma_2$ peak. In our model $R$ is the length of a uniform ionization density. It should be an overestimation of the straight line penetration distance, which is $\approx 3$ mm using the "practical range" formula of the PDG tables[6].

### 2.2.3 Determination of the ionization charge and the electron mean free path estimators

Decay rates and branching ratios of the sources have been checked independently with a solid state counter[3]. The measured values of the $\gamma_1$, $\gamma_2$ and $\gamma_3$ Compton edges are respectively 1, 2 and 5% below those predicted by kinematics. For $\gamma_1$ and $\gamma_2$, the width of the Compton edges is equal to 1.4 times the width of the corresponding electron conversion peaks, and for $\gamma_3$ it is 110 keV.

Aiming for a precise ionization charge measurement, we fit the upper edge of $\gamma_2$ with a straight line around 1 MeV. The energy window of this fit is defined by $2/3$ and $1/3$ of the peak height (see fig.4). On this line, we define as "$Q_{1\,MeV}$ edge" the point with $1/2$ of the peak height. In practice a statistical accuracy of $\approx 0.03\%$ (RMS, corresponding to 0.25 ADC units) is reached for this estimator within 5 minutes of data taking. Two other ionization charge estimators, using non-overlapping energy windows, have been defined: the "$Q_{482\,keV}$ peak" estimator (fit of a Gaussian peak plus a linear background around the $\gamma_1^K$ peak) and the "$Q_{\gamma_2}$ peak" estimator (fit of a parabola around the $\gamma_2$ peak).
The $S_{Bi}(Q)$ parametrization of $^{207}$Bi spectra provides the functions relating these 3 measured estimators and the mean free path $\lambda_e$. These functions are approximately equal to

$$Q(\lambda_e) = \lambda_e / D \times (1 - e^{-D/\lambda_e}) \times Q_0, \tag{1}$$

where $Q(\lambda_e)$ is the signal generated by an ideal charge $Q_0$ deposited on the cathode. This function $Q(\lambda_e)$ is derived by summing the elementary signal $dQ/dt$ generated by the drifting charge $Q_0$ (cf. [4]). It is often defined as a "*point-like*" charge collection efficiency [7][3], although it only implies that the ionization is deposited near the cathode.

We have tested the behavior of the $Q_{482\,keV}$ peak and $Q_{1\,MeV}$ edge estimators as a function of $\lambda_e$ by applying them to the $S_{Bi}(Q)$ parametrization of $^{207}$Bi spectra. For $\lambda_e/D > 0.4$, they follow within 1% the behavior of the function (1), as shown in reference[4] figure 3. In the following we shall apply this formula extensively to the 1 MeV charge estimator and define the inverse function $\zeta$ relating $Q$ to $\lambda_e$ by

$$D/\lambda_e = \zeta(Q/Q_0) \tag{2}$$

which can be approximated by

$$\zeta(Q/Q_0) \approx 2\log(Q_0/Q) \text{ for } \lambda_e/D > 2. \tag{3}$$

---

[3] thanks to G.Roubaud, Radio-protection group, CERN

## 3 Variation of ionization yield with temperature

This section is devoted to a "*liquid argon temperature effect*" (**LAr-T**) discovered in the course of our measurement. The analysis is based essentially on data recorded during a shutdown of HERA in January 1995. The temperature of the H1 LAr calorimeter was systematically modified. This "temperature scan" was sufficient to produce correction coefficients applicable to the analysis of the purity. However more experiments are needed to understand fully this LAr-T effect. They cannot be done during the normal operation of the H1 calorimeter.

### 3.1 LAr-T effect on the [207]Bi spectrum and the [241]Am spectrum

Figure 6 shows that the response of the "1 MeV" charge estimator follows closely the evolution of temperature in the LAr cryostat during the temperature scan of January 1995. We note that the two other estimators, "$\gamma_2$ peak" height and "$\gamma_1$ peak" (see § 2.2) display the same pattern with different systematic errors but larger statistical errors. It shows that the whole spectrum is shifted. This observation and a constant monitoring of the analog chain (cf. § 1.4 and § 1.5) both exclude the possibility of an artifact linked to an unexpected modification of the spectrum edge (e.g. electronic noise etc....).
In order to quantify the LAr-T effect  we made a linear fit$\Delta Q/Q = -\kappa\,\Delta T$ which yields $\kappa = $ *1.8%/K* for the 6 mm gap probe and *1.1 < $\kappa$ < 1.5%/K* for the 6 other probes with a 4 mm gap. We defined for this purpose a global cryostat temperature, average of 3 temperatures measured in 3 different regions of the cryostat (among the 7 ones which are constantly monitored). This procedure smooths out the rapid fluctuations of the local temperature due to the cooling method[4], which are too fast to be seen by the purity probes.

No LAR-T effect at all was seen on both [241]Am probes during the same period. This is clearly seen in figure 6 by comparing [241]Am with [207]Bi.

### 3.2 Noise introduced by thermal fluctuations

Our most precise purity measurement -the 1 MeV charge estimate- is affected by the thermal fluctuations of the H1 cryostat. Some fluctuations are large, reaching 0.5 to 1 K. They are mainly due to the delivery of liquid nitrogen from the factory twice per week (colder than the one used in the normal cooling circuit). They yield the tail of the distribution of $Q/<Q>$ above 2 $\sigma$ seen in figure 7 a).
Under normal conditions (sampling frequency $\approx$ 1 hour $^{-1}$), we see smaller thermal fluctuations which can be described as a noise with 0.025 < $\sigma_T$ < 0.05 K depending on the period of observation and the location of the probes as seen in figure 7 b). The noise affecting the [207]Bi "1 MeV" signal is consistent with a quadratic mean of the thermal fluctuation $\kappa\sigma_T$ and the 0.03% statistical error found in § 2.2.
The temperature map of the cryostat is not well known because the temperature probes are not calibrated absolutely. Most likely the temperatures does not vary beyond  0.7K. In this case extreme systematic effects affecting the purity measurements in various positions or at different times can reach the  1% level.

---

[4]  liquid nitrogen heat exchanger yields strong, non-reproducible, convection currents

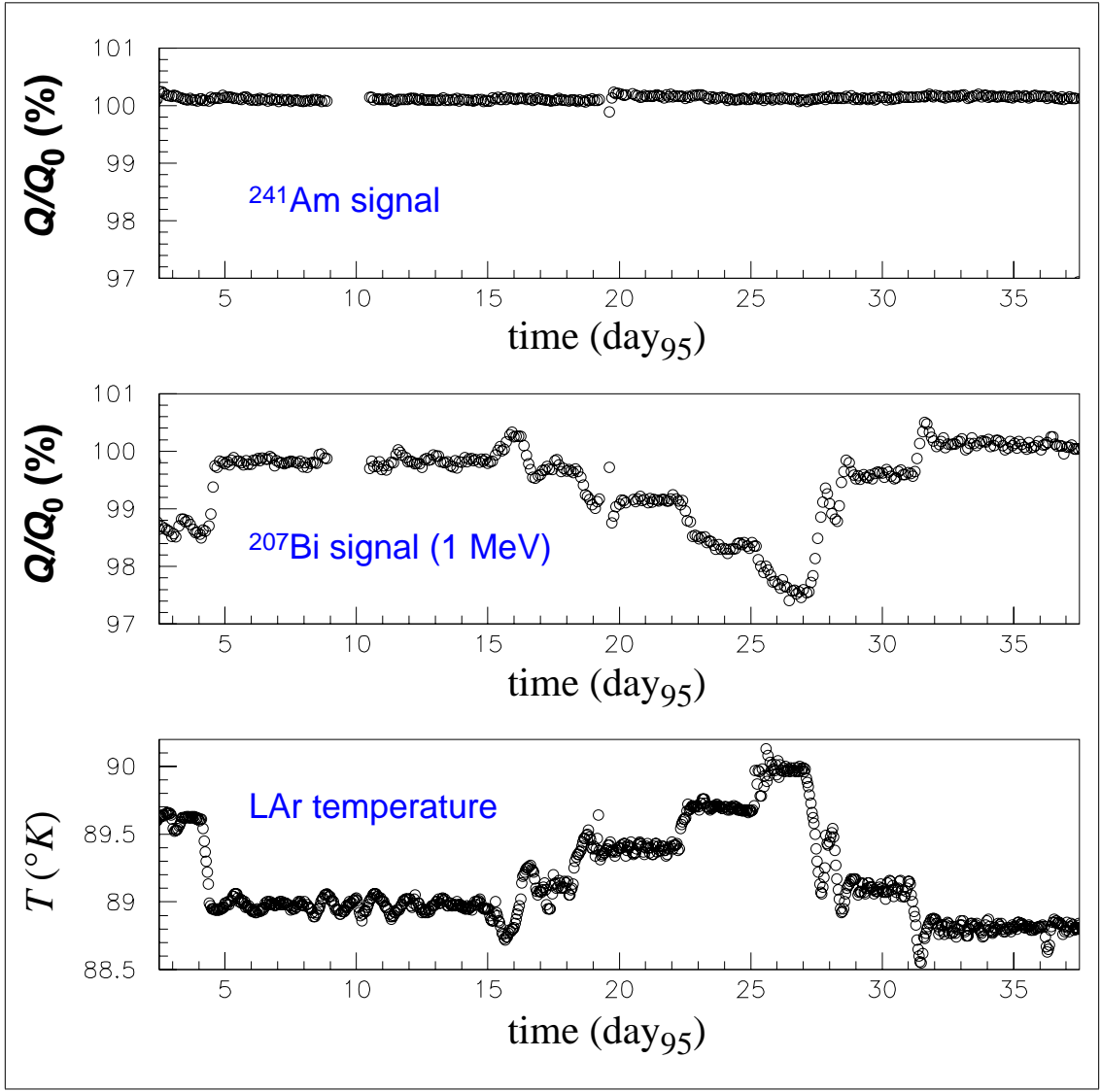

**Figure 6:** Relative variation of $^{241}$Am (probe 1, $E$=19.5kV/cm) and $^{207}$Bi (probe 2, $E$=6kV/cm) charge collection efficiency $Q/Q_{ref}$ (set to100% at day 9) during the liquid argon temperature scan (Dec. 94 to Feb. 95). The lower plot shows the global cryostat temperature, which is the average of 3 local measurements.

### 3.3 High voltage dependence of LAr-T effect

We make the hypothesis, based on the comparison of two high voltage curves taken at the same purity level with a "large" difference of temperature (1 K), that the temperature acts on the charge yield independently of the electrical field. We use this hypothesis in § 4 for a general analysis of the high voltage dependence.

### 3.4 Discussion of the liquid argon temperature effect

The LAr-T effect, about 1.5%/K, affects the $^{207}$Bi probes only. This excludes an electronic artifact which would affect equally the signals generated by each type of probe, because they have the same amplitude, the same shape and they are processed by the same readout chain. It eliminates also the effect of attachment of electrons crossing the gap. Such an effect should be visible with the [241]Am probe, although reduced in this case but not negligible. Quantitatively, the attachment effect on [241]Am probes has been measured to be a third of its effect on 6 mm [207]Bi probe during the oxygen leak in the H1 cryostat that occurred in 1994 and is described in § 5.3.2. This is clearly incompatible with figure 6.

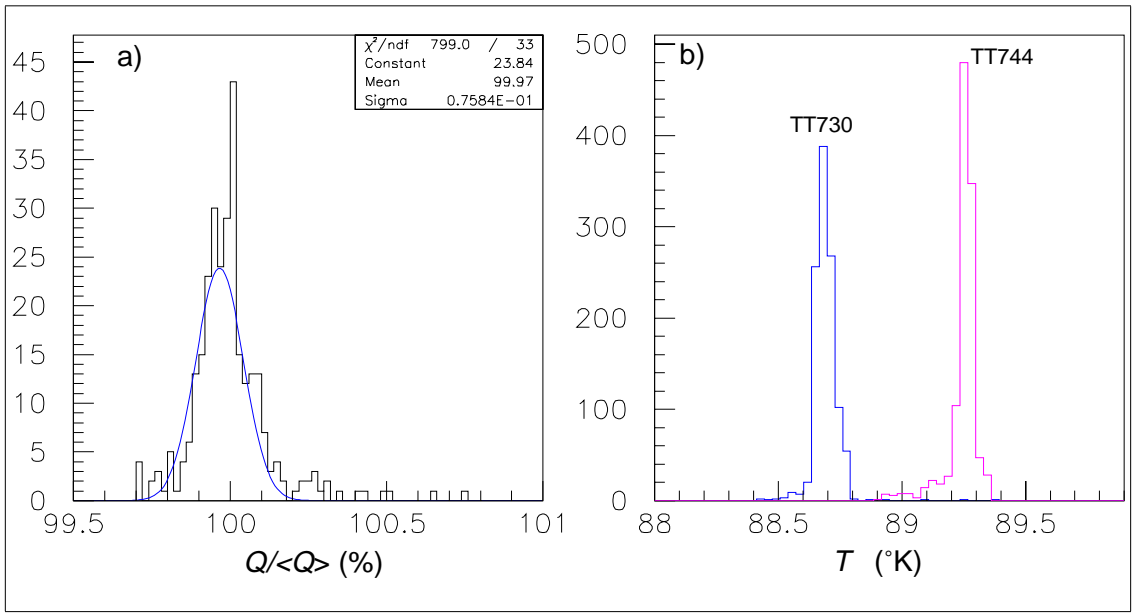

**Figure 7:** October 99: a) thermal fluctuation of purity ($Q/<Q>$ 6mm probe;1 MeV estimator); b) temperature distributions measured in 2 points of the cryostat (probes are not calibrated absolutely).

Two categories of explanations are left, based either on range or on ionization density. First let us consider a change of range. We would expect a much smaller effect on the 482 keV than on the 1 MeV line contrary to the facts. Moreover the range dependence is governed by the density variation (0.6% per K) which displaces the center of gravity of the ionization towards the cathode. In our model of section § 2.2 this reduces by 0.2% the charge collection efficiency of 1 MeV tracks perpendicular to the cathode (worst case) and not at all the "point-like" tracks, i.e. the 482 keV and 1 MeV estimators. Another indirect effect of range is the modification of the electric field due to the space charge of slowly drifting Ar+ ions in case of β-sources. The corresponding voltage drop, taking an ion/electron drift speed ratio ~10[-5] and a β-source activity ~1 GeV/s, would be a few volts over 4 kV, with a minimal effect on the ionization yield.

Second let us examine the recombination effects related to the high density of ionisation of α's. According to the classical models[9][11], recombination decreases with increasing mobility for which a temperature dependence of -1.7%/K has been measured recently[10]. Decreasing temperatures should therefore enhance the yield of α-sources (75% recombination) much more than the yield of β-sources (1% recombination), contrary to what we see.

In summary, we cannot explain the absence of LAr-T effect with alphas as well as its presence with betas, with the classical recombination and ionization yield models.

We cannot say presently if the LAr-T effect is specific to our purity probes or if it may be generalized to the H1 calorimeter. More information from other calorimeter tests with high energy beams is needed.

## 4 High Voltage Curves for the $^{207}$Bi Probes

The electron mean free path $\lambda_e$ and purity can be related by the assumption that attachment is a simple process with a cross-section:

$$\sigma(E) \ = \ 1/(\lambda_e(E)N) \qquad\qquad (4)$$

where $N$ is the number of impurities per unit of volume and $E$ the electrical field.

Since the work of [7] $\lambda_e$ is usually assumed to be proportional to $E$ ("$\lambda \propto E\ model$"). For electron sources this model assumes in addition that the ionization yield $Q_0$ does not depend on $E$. Following Thomas and Imel[11], some authors[3][12], look for a better representation of electron source data by introducing an $E$ dependence of $Q_0$ ("$box\ model$" of electron-ion recombination). Recombination, in electron sources, does not alter significantly the shape of HV-curves[8]. Its effect is equivalent to an increased LAr impurity. This is demonstrated in appendix 2 of reference[4].

In this section we shall use the "high voltage curves" relating the charge $Q$ collected in an ionization chamber to the electric field $E$ in order to determine $\sigma(E)$ in a model independent way. Then we shall determine the impurity level $N$ from measured high voltage curves.

### 4.1 Determination of $Q_0$ and $\sigma(E)$

The analysis presented here yields values of $\sigma(E)$ over a large range of field strengths ($.5 < E < 15$ kV/cm) and the asymptotic ($E \rightarrow \infty$) ionization yield $Q_0$. As $Q_0$ can be defined for each probe in a model independent way with a statistical precision better than 0.2% (RMS), we refer to it as calibration constant of the probe. It is based on a sequence of high voltage curves, taken periodically between 1991 and 1998 on two $^{207}$Bi probes (with respectively 4 and 6 mm gaps). Although the impurity concentration $N$ in the H1 cryostat increased during this period and the temperature of the cryostat varied within an interval of 1.5 $^0$K, the purity and the temperature were stable during the taking of each high voltage curve, because of the inertia of the large volume of liquid argon (56 m$^3$).

We have encountered two main problems:

- At high field the variation of $Q(E)$ with $N$ is very small. We only can profit from the 0.03 % relative precision of the 1 MeV charge estimator, if we are able to control the temperature induced charge variations $\Delta Q_0/Q_0 = -\kappa\,\Delta T$. This is done using a HV-curve averaging method (see below).

- At low field, where the maximum drift time is of the order of the shaping time of the electronics, the conditions of a pure charge integration may not be met anymore. We studied the effect of different time constants on the HV-curves and showed that a 3 μs shaping time eq.(1) can be applied for $E > 1$ kV/cm. This will be checked experimentally in § 5.1.

On the basis of eqs. (2) and (4) we extract $N\sigma$ from

$$DN\sigma(E) = \ D/\lambda_e(E) = \ \zeta(Q(E)/Q_0)\,. \qquad\qquad (5)$$

A measured HV-curve can be represented in the ($E^{-1}$, $D/\lambda_e$) plane using this relation with an approximate value for $Q_0$, as schematically shown in figure 8. $D/\lambda_e$ is proportional to the impurity $N$ and two curves are drawn for some impurities $N_l$ and $N_h$.

A shift $\Delta Q_0$ of the calibration constant yields approximately a translation parallel to the $D/\lambda_e$ axis. (cf. the approximation of $\zeta$ in equation (3)). We therefore calibrate the probe by a shift $\Delta Q_0$ which makes $D/\lambda_e$ and $N_l$ proportional for different fields. For that purpose we compare in practice the resulting $D/\lambda_e$ values for the two field values $E_1$ and $E_2$ at the two impurities $N_{,l}$ and $N_{,h}$ as sketched in figure 8.

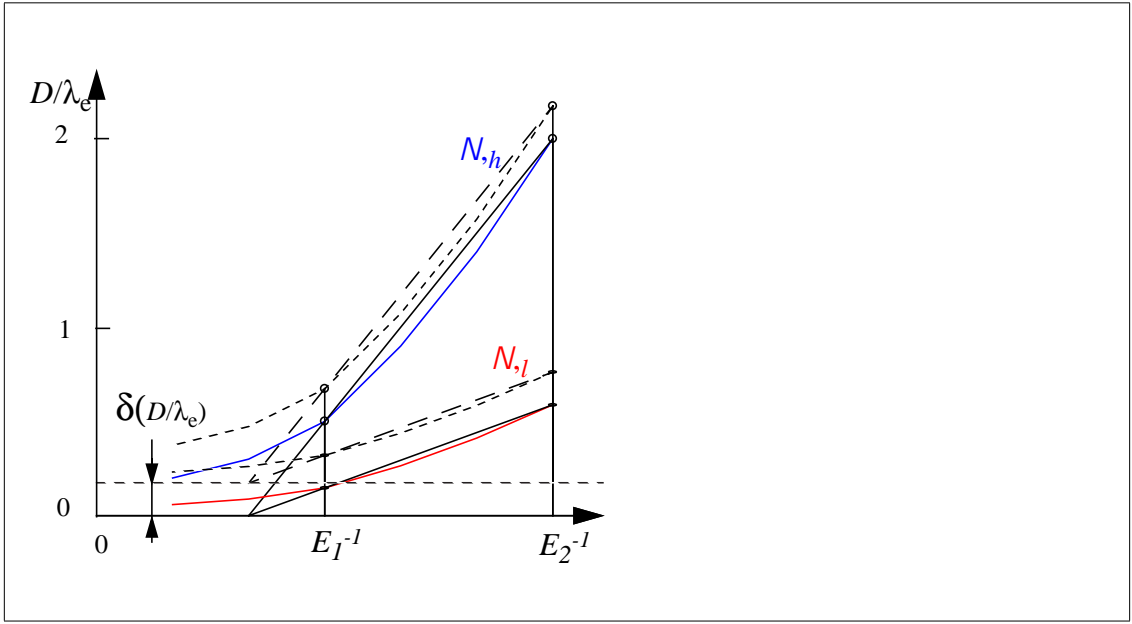

**Figure 8:** Properties of HV-curves in the ($E^{-1}$, $D/\lambda_e$) plane (schematically): 1) changing the purity from $N_{,h}$ to $N_{,l}$ changes the $D/\lambda_e$ scale by a factor $N_l/N_h$; 2) changing the calibration constant from $Q_0$ (solid) to $Q_0 + \Delta Q_0$ (dashed) translates curves by $\delta(D/\lambda_e)$. The straight lines illustrate the calibration procedure.

We obtained the calibration constants for both probes (4 and 6 mm gap) by applying this calibration procedure not to single curves at different impurities but instead to an average over a set of curves taken at low impurities in the years 1991-93 and another average over a set of curves taken at high impurities in the years 1993-98. The impurities of the two sets differ by $N_h/N_l \approx 3$.

After independent calibration of the two probes, we compared the resulting $\langle DN_h\sigma(E)\rangle = \langle D/\lambda_e(E)\rangle$ and found excellent agreement with the expected ratio of $1.5 = 6$ mm/ 4 mm as shown in fig. 9. For this comparison we used the 1993-98 data. Here we have a set of 20 HV-curves taken for both probes at the same time[5]. The higher impurities of this period lead to a larger cross section to noise ratio. This comparison of two different probes at the same field constitutes a strong check of systematic errors affecting the absolute measurement of $\lambda_e$.

The resulting 1993-98 averages <$1/\lambda_e(E)$> for both probes are shown in figure 10 . Here we have added four high field points above $E$= 10 kV/cm measured in year 2000[6]. The behavior of $1/\lambda_e(E^1)$ over the whole range of $E^1$ is reproduced empirically by the formula

---

[5]  When necessary we interpolated the 4 mm probe data linearly to the $E$-field values of the 6 mm probe data

[6]  Impurity $N$ and yield $Q_0$ were artificially restored to their average 1993-98 values using the linear transformations described in fig.8 and the field dependence at $E$<10 kV/cm of the year 2000 HV-curve.

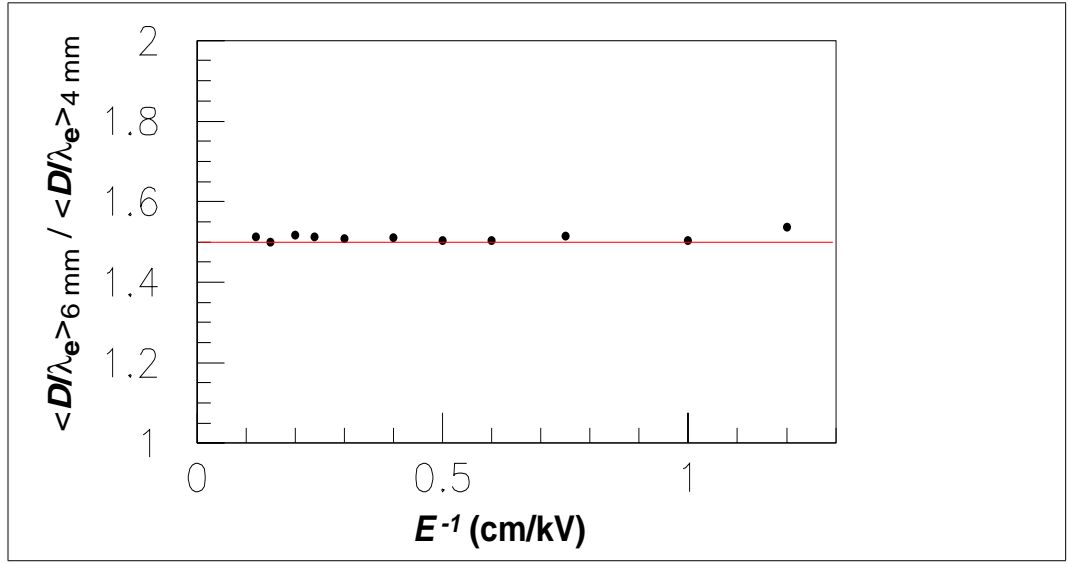

**Figure 9:** The ratio of relative absorption length in 6 and 4 mm probes as a function of $E^{-1}$. The horizontal line represents the gap ratio. The averages $<D/\lambda_e>$ are taken over all 1993-98 data.

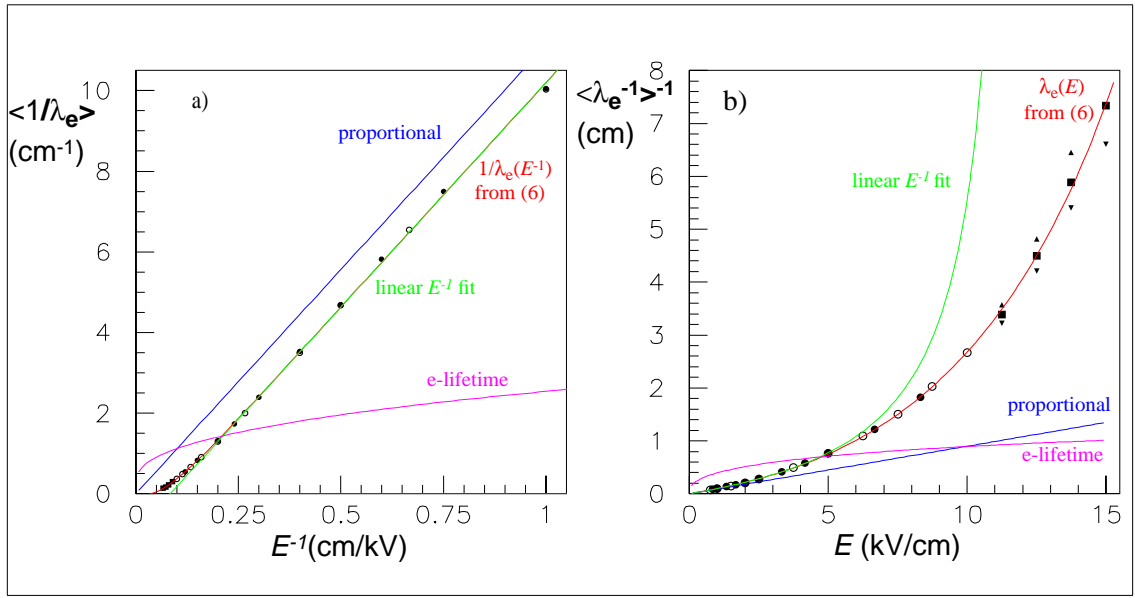

**Figure 10:** $<1/\lambda_e>$ vs. $E^{-1}$ or $<\lambda_e^{-1}>^{-1}$ vs. $E$ averaging 20 high voltage curves taken in the years 1993-98 for the $^{207}$Bi probes with 4 mm (○) or 6 mm (●) gaps and extra high field points (year 2000; 4 mm ■). The LAr-T effect is indicated at high field ($^+_-$ 0.3K symbols: ▲▼). Four attachment models are compared to the data. The empirical fit of $\lambda_e(E)$ from eq.(6) (solid line), the modell[7] where $\lambda_e \propto E$ (dashed), a linear $E^{-1}$ fit (dotted), and the electron lifetime model $\lambda_e \propto v(E)$ (dashed-dotted).

$$1/\lambda_e(E) = 1/\lambda_0 \times \left( E_S E^{-1} + e^{-E_a E^{-1}} - \eta \right). \qquad (6)$$

Figure 10 shows that this formula gives an excellent fit between $E= 1$ and 15 kV/cm. Fitted values are $E_S$=12.0 kV/cm, $E_a$=15.7 kV/cm, $\lambda_0$=1.076 cm, $\eta$=1.004. The fact that for $E^{-1} \rightarrow 0$ the values of $1/\lambda_e(E^{-1})$ and its derivative are respectively much smaller than for e.g. $E^{-1} \approx 0.2$ cm/kV yields the two following constraints: $\eta \approx 1$ and $|E_a - E_S| \ll E_S$. The linearity of $1/\lambda_e(E^{-1})$ for $E^{-1} > 2/E_s$ is built in the formula.

Figure 10 also shows that the classical *"electron lifetime model"*[7], in which $\lambda_e(E)$ is proportional to $v(E)$, is incompatible with the data because $\lambda_e(E)$ and $v(E)$ have opposite sign curvatures.

## 4.2 Determination of an impurity scale

Our ansatz is that the electron mean free path $\lambda_e$ is always given as a function of $E^{-1}$ by (6), with only one parameter, $\lambda_0$, depending on purity. We define two impurity scales based either on an effective oxygen contamination or on the observed change of response of the H1 electromagnetic calorimeter (<u>*EM*-calo</u>) to electrons.

### 4.2.1 Oxygen impurity scale

The "*equivalent concentration of oxygen*" $N_{O_2}(ppm[O_2])$, producing a given $\lambda_e$, is based conventionally on a model where $\lambda_e$ is proportional to $E$. It is parametrized by:

$1/\lambda_{O_2}(E) = N_{O_2}\sigma_{O_2}$, with $\sigma_{O_2}(E) = E^{-1}/0.15$ $(cm^{-1} \cdot ppm^{-1})$ for $E < 5$ kV/cm[8].

Looking at figure 10a, we see that this model cannot fit the data. But the data points with $E < 5$ kV/cm can be described by a linear fit with a good $\chi^2$ (as obtained also in ref.[7]). Such a fit leads to a resulting calibration constant $Q_0$ which is not independent of the purity in contrast to its definition (see § 4.1). If instead we modify $\sigma_{O_2}$, consistent with eq.(6), to:

$$\sigma_{O_2}(E) = \left(E_S E^{-1} + e^{-E_a E^{-1}} - \eta\right)/0.15 E_S \quad (cm^{-1} \cdot ppm^{-1}) \qquad (7)$$

then $N_{O_2}\sigma_{O_2}$ fits the 1993-98 average $<1/\lambda_e>$ with $N_{O_2} = 0.15 E_S/\lambda_0 = 1.67\,ppm[O_2]$.

### 4.2.2 Impurity scale of the H1 *EM*-calo

We define an impurity scale $N_{H1}$ by the relative decrease $\Delta Q/Q_0$ of collected charge in the H1 *EM*-calo. We further introduce the effective cross section $\sigma_{H1}$ with $N_{H1}\sigma_{H1}(E)=1/\lambda(E)$. In the $\lambda \to \infty$ approximation in Table 1, we obtain $\qquad 1/\lambda(E) = (3/D)\,\Delta Q/Q_0$.

**Table 1:** Parameters of H1 liquid argon cells at DESY

| source | [241]Am probe | [207]Bi probe | [207]Bi probe | *EM*-calo |
|---|---|---|---|---|
| gap $D$ (cm) | 0.2 | 0.4 | 0.6 | 0.235 |
| $E_O$ (kV/cm) | 19.5 | 6.5 | 6.0 | 6.38 |
| [a]$log(Q_0/Q)^{(\lambda\to\infty)}\approx$ | $D/2\lambda$ | $D/2\lambda$ | $D/2\lambda$ | $D/3\lambda$ |
| $\sigma(E_O)/\sigma(6.38\text{ kV/cm})$ | [b]0.8±0.25 | 0.97 | 1.11 | 1 |
| $G$ (sensitivity factor) | [b]1.15±0.3 | [c]2.48 | [d]4.25 | 1 |

a. basic formulas and 1st order $1/\lambda$ approximations are found in appendix 1 of [4]

b. $G$ is measured as the ratio of relative signal variation of Am and Bi probes during the air leak event (Bi variation expressed in H1 scale) and then used to determine $\sigma(E_O)/\sigma(6.38$ kV/cm)

c. computed by $G=(3/2)$ x $(D/0.235$ cm) x $(\sigma(E_O)/\sigma(6.38$ kV/cm)), see eq.(9)

---

[7] In this model the lifetime $\tau=\lambda_e(E)/v(E)$ is a constant; the drift speed $v(E)$ is taken from reference [10].

[8] The coefficient 0.15 ppm·cm²/kV is taken from ref. [7]. In ref. [3] it is found to be 0.088 ppm·cm²/kV.

that is with the *EM*-calo parameters (Table 1), $\sigma_{H1}$(*6.38 kV/cm*)= 3/D= 12.77 *cm$^{-1}$*.

The two impurity scales are related by $N_{O_2}\sigma_{O_2} = N_{H1}\sigma_{H1} = 1/\lambda_e$.

At *E=6.38* kV/cm, since from (7) we have $\sigma_{O_2(6.38kV/cm)} = 0.53 \quad (cm^{-1}\cdot ppm^{-1})$,

this relation reads $N_{O_2} = N_{H1}\cdot 12.77/0.53 = N_{H1}\cdot 24.1(ppm[O_2])$,

that is a signal loss of 1% in the H1 *EM*-calo ($N_{H1}$= 1%) corresponds to an increase of 0.24 ppm[$O_2$] in the oxygen impurity scale.

### 4.2.3  Impurity estimator

Any point-like charge collection estimator $Q$ at a field $E$ can be converted (eqs. (2) and (4)) into an impurity estimator (including the LAr-T correction $\kappa\Delta T$):

$$N_{H1} = \zeta((1 + \kappa\Delta T)\cdot Q/Q_0)/D\sigma_{H1}(E). \tag{8}$$

## 5  Impurity Data

### 5.1  Data collected at HERA

A first application of the impurity estimator (8) consists of converting each HV-curve into an impurity curve using the electric field dependence of (6). The basic law (4), which implies that $N$ does not vary with $E$, is well verified for the individual curves: the point to point variations in the medium or the high field regions of the curves are of the order 0.1% (RMS) (see fig. 11).

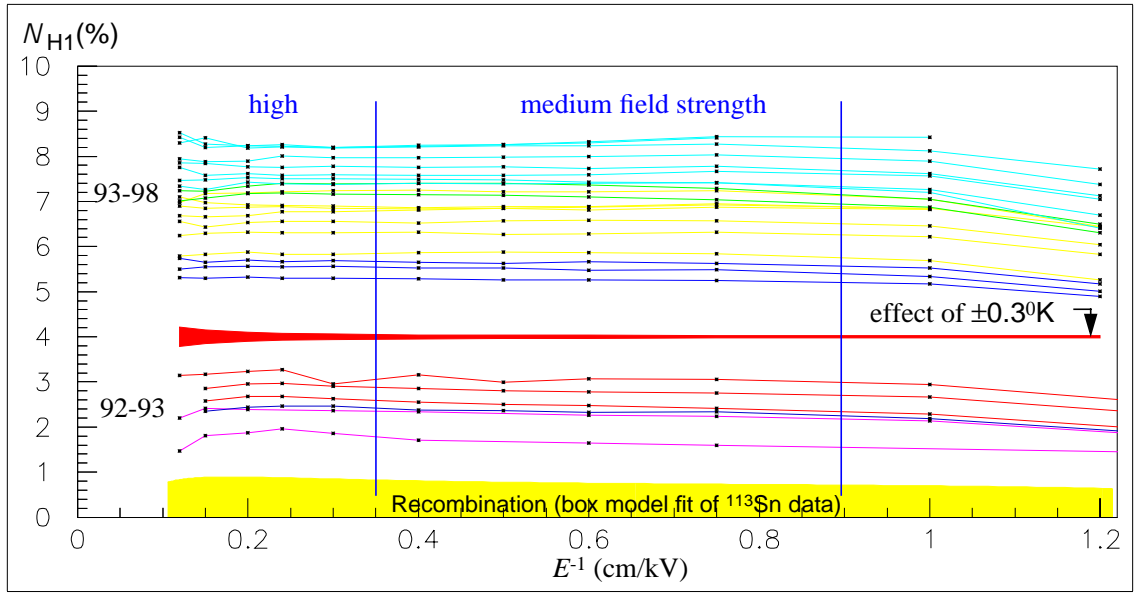

**Figure 11:**  HV curves converted into impurity curves for the 6 mm probe (1991-98). The medium field region defines an average impurity $\overline{N}$ for the monitoring of the H1 calorimeter over the years. The part of the effective impurity due to recombination, according to a "box model"[11], is represented as a band above the horizontal axis.

The main effects violating this law are:

a) for E > 3 kV/cm, a temperature uncertainty. (The effect of a ±0.3 K shift around $N_{H1}$= 4% is shown in fig. 11).

b) for E < 1 kV/cm, a large absorption (particularly for a 6 mm gap) and a drift time not negligible compared to the integration time of the readout electronics.

c) for $N_{H1}$< 2%, a contribution to the effective impurity due to recombination[9].

For each impurity curve an average impurity $\overline{N}$ is obtained for the field range $0.35 < E^{-1} < 0.9$ cm/kV. The difference between the $\overline{N}$ measurements from the 4 and the 6mm probes (shown in figure 12) is below 0.2% (RMS).

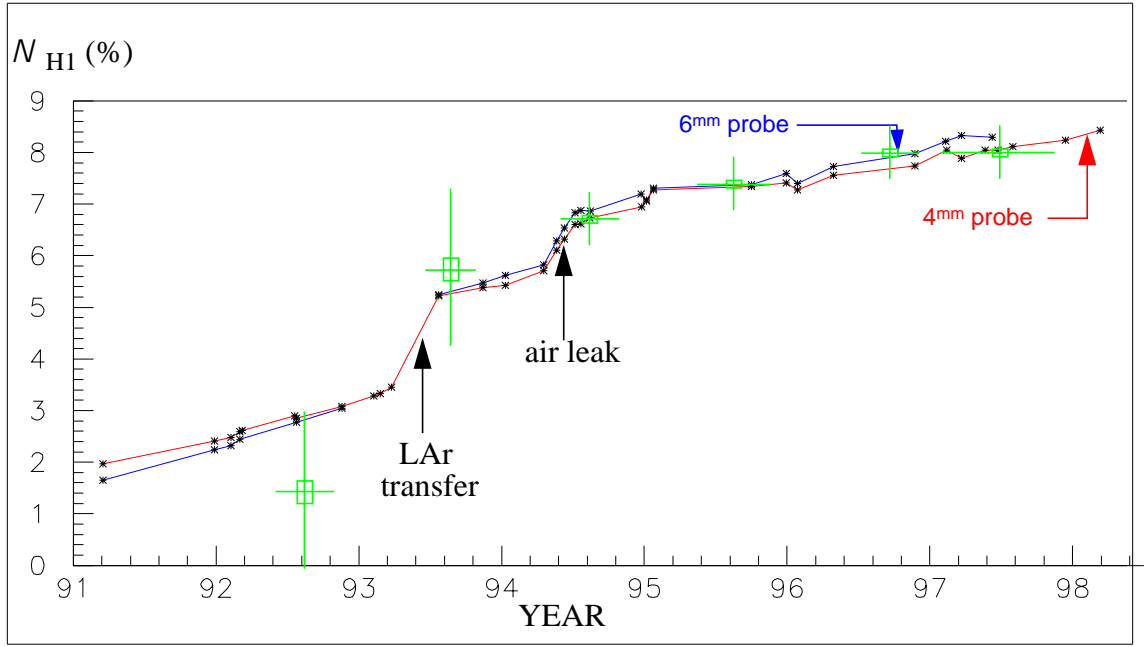

**Figure 12:** Impurity increase in H1 from 91 to 98 seen by 4 and 6 mm probes. Also shown is the time variation of overall correction factors as determined by kinematic calibration of H1 *e-p* events with respect to the known electron beam energy (arbitrarily set to 9% for 1998).

A second application of the impurity estimator is the conversion of the monitoring signals (1 MeV charge estimator in case of $^{207}$Bi) into an impurity measurement as shown in figure 13. However it requires the knowledge of the calibration constant $Q_0$ coming from the fit of one HV-curve (often missing for the beam tests discussed later). But, regardless of $Q_0$, measured charge variations in the probes can, by differentiating eq. (8) and using eq. (3), be related to impurity variations in the H1 calorimeter by:

$$G\Delta N_{H1} = -\Delta Q/Q - \kappa\Delta T \qquad (9)$$

Here $G$ is the ratio of the impurity sensitivity of a probe with a field $E_O$ and the H1 *EM*-calo with field 6.38 kV/cm. $G$ is the product of 3 terms: the drift distance ratio, the impurity cross section ratio for the different fields and a factor 3/2 taking into account the effect of an extended source compared to a point source.

The raw monitoring data in figure 13, are normalized by the factor $G$, but they lack a point to point temperature correction.

---

[9]  It is expected to be ≈0.7% for a 6 mm probe as indicated in fig.11 using the box model of [1.1] fitted on $^{113}$Sn data[13], but is expected to be smaller for $^{207}$Bi.

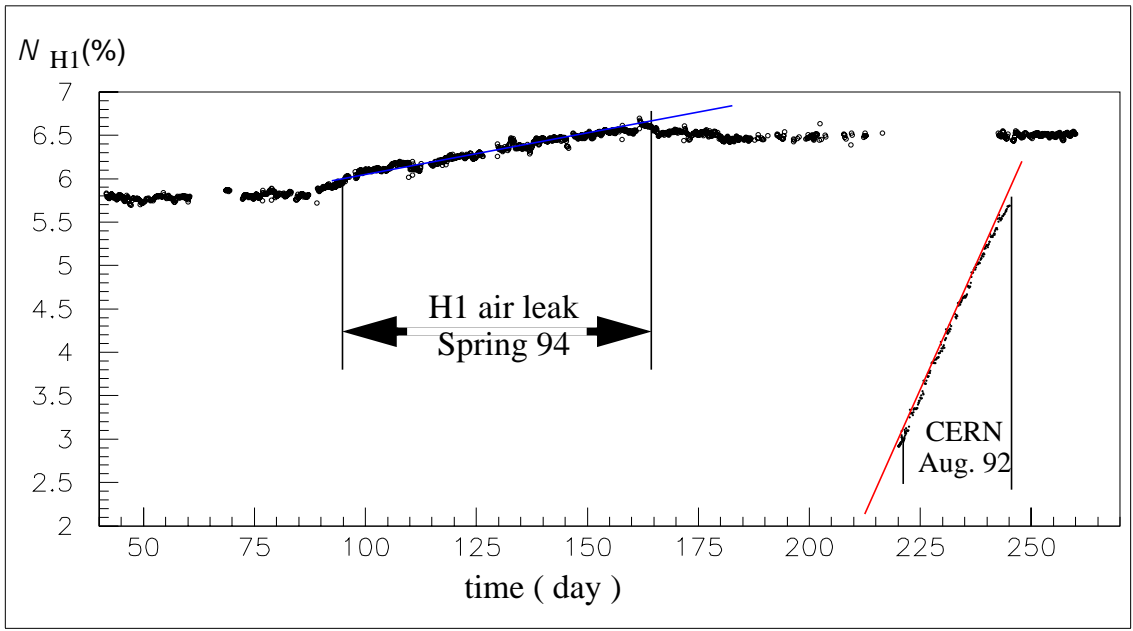

**Figure 13:** Impurity monitoring with $^{207}$Bi 6mm probes at $E$=6kV/cm (○ H1 94; • CERN 92).

## 5.2 CERN test beam data

During the years 1988 to 1992 each type of the H1 calorimeter modules was installed in a cryostat[14] located in the H6 test beam at CERN. The purity monitoring system used for the H1 LAr calorimeter was developed during these tests. In 1992, during the last two running periods, it consisted of four $^{207}$Bi and two $^{241}$Am probes described in Table 2 functioning like the H1 system installed at DESY in January 91. Earlier the system at CERN contained only

**Table 2:** Parameters of H1 liquid argon cells at CERN (August 92)

| source | $^{241}$Am probe | $^{207}$Bi probes | | | | *EM* calo |
|---|---|---|---|---|---|---|
| gap $D$ (cm) | 0.2 | 0.4 | 0.4 | 0.6 | 0.6 | 0.235 |
| field $E_O$ (kV/cm) | 20.0 | 12. | 10. | 8.33 | 6.66 | 10.64 |
| $log(Q_0/Q)^{(\lambda \to \infty)} \approx$ | $D/2\lambda$ | $D/2\lambda$ | | | | $D/3\lambda$ |
| $\sigma(E_O)/\sigma(6.38 \text{ kV/cm})$ | [a]0.63±0.12 | 0.50 | 0.58 | 0.78 | 0.96 | 0.58 |
| $G$ (sensitivity factor) | [b]0.81±0.15 | [b]1.3 | [b]1.5 | [b]2.9 | 3.7 | [b]0.53±0.05 |

a. deduced from the corresponding sensitivity factor
b. measured as in Table 1 using signal decrease during 25 days, relatively to the Bi probe with 6.66 kV/cm

two $^{207}$Bi probes and its purity data were less precise and reliable.

During a test period of 2 to 6 weeks, the calorimeter and the probe signals decreased due to a continuous release of impurities into the LAr. A decrease in the signal of a few percent per month was observed, as shown in figure 13 for the August 92 run.

For this run a sensitivity factor $G$=0.53±0.05 was measured for the H1 test module by comparison with the signal decrease in a $^{207}$Bi probe ($E_O$=6.66 kV/cm). This is to be compared with the value $G$=0.58 expected from the impurity cross section ratio given in Table 2 for the different fields (10.6 vs. 6.38 kV/cm, since the nominal drift field was reduced in 1993). Therefore the attachment cross sections seen by the *EM*-calo and the $^{207}$Bi probes are the same within experimental errors.

Let us remark that the effective cross section $\sigma(E_O)$ measured with $^{241}$Am sources around 20kV/cm, either in the H1 calorimeter or in the test modules at CERN (Table 1 and 2), are roughly compatible. However the ratio of effective cross sections, $\sigma(20\text{kV})/\sigma(6.38\text{kV})$, does not correspond to the observed field dependence of the Bi probes (cf. fig.10b), indicating that for $^{241}$Am the effective cross section is not simply due to attachment.

The rise of impurity during each CERN run has been monitored by various probes and was eventually compared to *EM*-calo modules, using the G factors of Table 2. As seen in fig.14, these rises are consistent within any one run but vary wildly from run to run. In this context of problematic LAr pollutions, as discussed in the next section, the role of $^{207}$Bi impurity probes is essential.

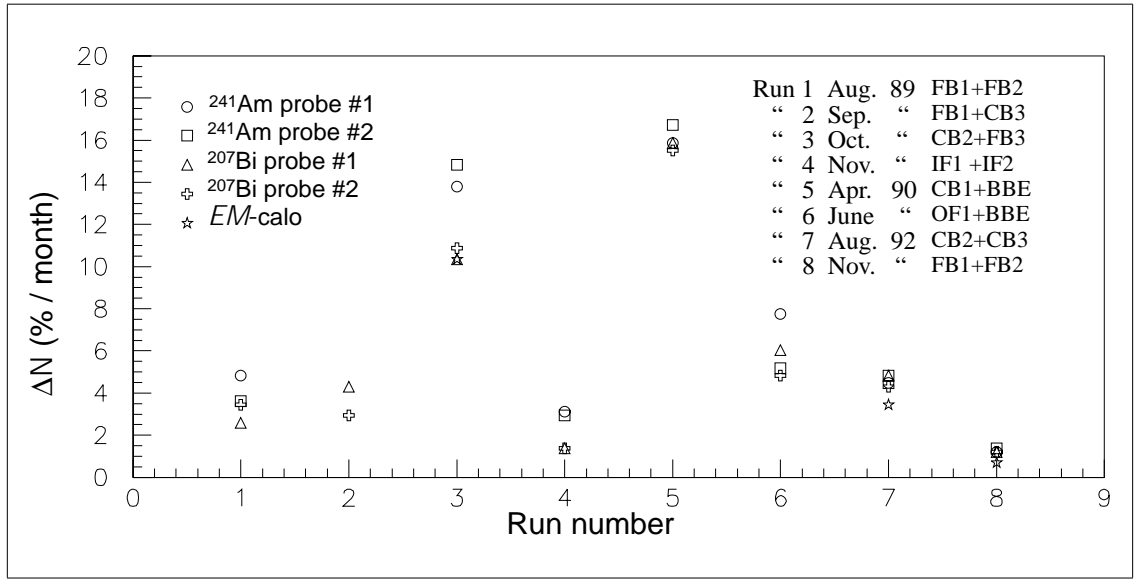

**Figure 14:** Rise of impurity during the tests performed at CERN (% / month) vs. run number (each run is characterized by its date and calorimeter module IDs).

### 5.3 Pollution monitoring throughout the operation of the H1 calorimeter

#### 5.3.1 CERN operations 1988-92

For each CERN run the load of LAr was renewed. This required the measurement of the purity level at delivery and its evolution throughout the duration of the test, as explained in § 5.2. During these calibration runs we have observed different levels of degradation of the argon purity with time for different types of calorimeter modules [15]. Other facts indicate that the pollution of the LAr was caused by the calorimeter modules and not by the cryostat or common accessories (cables, connectors,...), although only one clear correlation with a manufacturing procedure could be established (some elements of the modules of the CBE family were cleaned with freon FLUGENE). In particular, a comparison of the data produced by various probes could locate the pollution sources when the argon was kept motionless (by stratification of LAr inside the cryostat). The flow of pollutant was always constant, as shown by the time dependence in fig.13 for one of the runs. *A posteriori* chemical analysis, supported by results from the on-line oxygen detector [10], proved that the pollutant was not oxygen at the ppm level but some of the non identified molecules present at a few ppb level. There has been no sign of self cleaning of a CBE module, tested in 1992 after more that one year of exposition to air at room temperature.

### 5.3.2  DESY H1 operations 1992-1997

Figure 12 shows the time variation of the signals recorded since the beginning of the experiment in early 1991.The system was kept in operation throughout the periods where H1 was taking data. All data are stored in a dedicated database and monitored through a WWW page.

The important loss of signal in 1993 can be explained by the introduction of impurities present in the transfer lines which were used during the emptying and refilling operation when moving the detector. After 1994 the cryostat was moved full, therefore this source of contamination was significantly reduced. The rapid decrease of the signal observed during Spring 94 was caused by a slight leak in the valve between the cryostat and the eduction line. This leak was detected thanks to the constant monitoring of the probe signals (cf. fig.13) but it took several weeks to locate and fix it.

The results of over six years of operation show that cumulative effects of the liquid argon pollution on H1 energy measurement are around 2.5%, when not taking into account the contamination brought by the liquid argon transfer operations and the leak event. This is 100 times less than the pollution rate seen at CERN in August 92 (or 500 times less than in April 90).

This huge difference of pollution between CERN and H1 can be attributed either to a polluting piece not present in the final H1 assembly or more likely to the cleaning procedure used at DESY. The CERN procedure, based on pumping and outgassing the detector in a secondary vacuum, was not feasible with the H1 cryostat. The DESY flushing and cool down procedure was different. Flushing was done by circulating warm gas through this loop and pumping out the cryostat (3 times). The cryostat was set up in a loop with an external liquid nitrogen heat exchanger and cooled down by circulation of helium gas. After about 30 days, once liquid argon temperature was reached, the cryostat was filled with liquid argon (2 days). A possible interpretation of the fact that the high rate of pollution observed for the CERN tests was not seen any more could be that the surface of the heat exchanger acted as an impurity trap. The DESY cleaning procedure was tried in our last CERN test in 1992, without success. This confirms that the flushing and cool down procedure used in the H1 cryostat in DESY was much more efficient mainly because the circulation of gas was forced through the detector, lasting about a month, in comparison to the few days of the CERN tests.

## 6  General conclusion

We are able to measure $\lambda_e$ with our [207]Bi probes very precisely. This precision allowed us to detect electro-negative impurities diffusing in the LAr, coming for instance from minute air leaks.

We observe a temperature dependence for the charge collected in $\beta$ but not in $\alpha$ probes.

---

[10] SIMAC DS2500 oxygen analyzer

There is no explanation for this.

From the monitoring of the LAr purity two kinds of information were obtained for the H1 calorimeter. First, the long term variation of the calorimeter response due to pollution is well under control. Knowing the real electron attachment cross-section instead of approximate models, we were able to show that the long term purity measurement was consistent within ≈0.2% RMS for two different probes, after correction of thermal fluctuations of order ≤1%. For physics analysis corrections are needed when combining the data taken in various times. However we do not know the exact sensitivity of the H1 *EM*-calo to temperature which could be similar to that of our probes. In this case the spatial fluctuations of the calorimetric response due to temperature gradients could be in the 2% range. In practice local calibration factors are used for physics analysis which are obtained exploiting the kinematic constraints of HERA *e-p* events[16].

The observed temperature effects may be of interest for high precision calorimetry and more generally for the understanding of electron transport and recombination in LAr.